# Environmental Atmospheric Turbulence at Florence Airport

Salvo Rizzo[1,2] and Andrea Rapisarda [2]

[1] *ENAV S.p.A. U.A.A.V Firenze,*
[2] *Dipartimento di Fisica e Astronomia and Infn sezione di Catania, Università di Catania, CACTUS Group: www.ct.infn.it/~cactus*

**Abstract.** We present an analysis of a time series of wind strength measurements recorded at Florence airport in the period October 2002 – March 2003. The data were taken simultaneously by two runway head anemometers, located at a distance of 900 m, at a frequency of $3.3\cdot10^{-3}$Hz. The data show strong correlations over long time spans of a few tens of hours. We performed an analysis of wind velocity as it is usually done for turbulence laboratory experiments. Wind velocity *returns* and wind velocity differences were considered. The pdfs of these quantities exhibit strong non-Gaussian fat tails. The distribution of the standard deviations of the fluctuations can be successfully reproduced by a Gamma distribution while the Log-normal one fails completely. Following Beck and Cohen *superstatistics* approach, we extract the Tsallis entropic index q from this Gamma distribution. The corresponding q-exponential curves reproduce with a very good accuracy the pdfs of returns and velocity differences.

## INTRODUCTION

During the last decade enormous effort has been devoted to understand the physical origin of turbulence, performing various experiments with controlled flows. The Kolmogorov hypotheses have been verified in the laboratory as well as in the atmospheric boundary layer. For the latter, the measurement has to be made sampling the flow velocity for a relative short time and with a high sampling rate and for relatively constant mean velocity flow in order to control and maintain constant the Reynolds number. In our experiment we cannot control the Reynolds number since our measurements, taken at Florence airport, were done for a time interval of six months (from October 2002 to March 2003). Data were taken by using two runway heads anemometers, located at a distance of 900 m and with a sampling frequency of one sample every 5 min. However, we have found several features of "canonical" turbulence as we will show in the following. The aim of our study was to perform a an analysis of our data with the usual mathematical tools adopted in hydrodynamics turbulence and to reproduce the pdfs of the intermittent velocity components. In particular, this was done following the nonextensive approach adopted in refs. [1-4], and the concept of *superstatistics* introduced in ref. [5], which justifies the successful application of *Tsallis statistics* in different fields, and more specifically in turbulence experiments [1-4,6].

In this paper we report only a short description of the experiment and of the analysis we performed. A more detailed and complete discussion will be published elsewhere [7].

## THE EXPERIMENT

Wind velocity was recorded at Florence airport (43°48'35N 11°12'14"E) by two analogical anemometers, each one mounted on a 10 m high pole, located at the two runway thresholds and referenced in the text with "RWY05" and "RWY23". The height of the two anemometers were respectively 37,49 m and 40,23 m above mean sea level. The measures were "canalized" taking the nearest whole wind module (in Knots: 1Kts ≈1.82 Km/h ) and one of 36 angle windows, for the wind direction (tenths of degrees). With those measures we have derived the time series of the longitudinal and transversal components with respect to the runway principal axis.

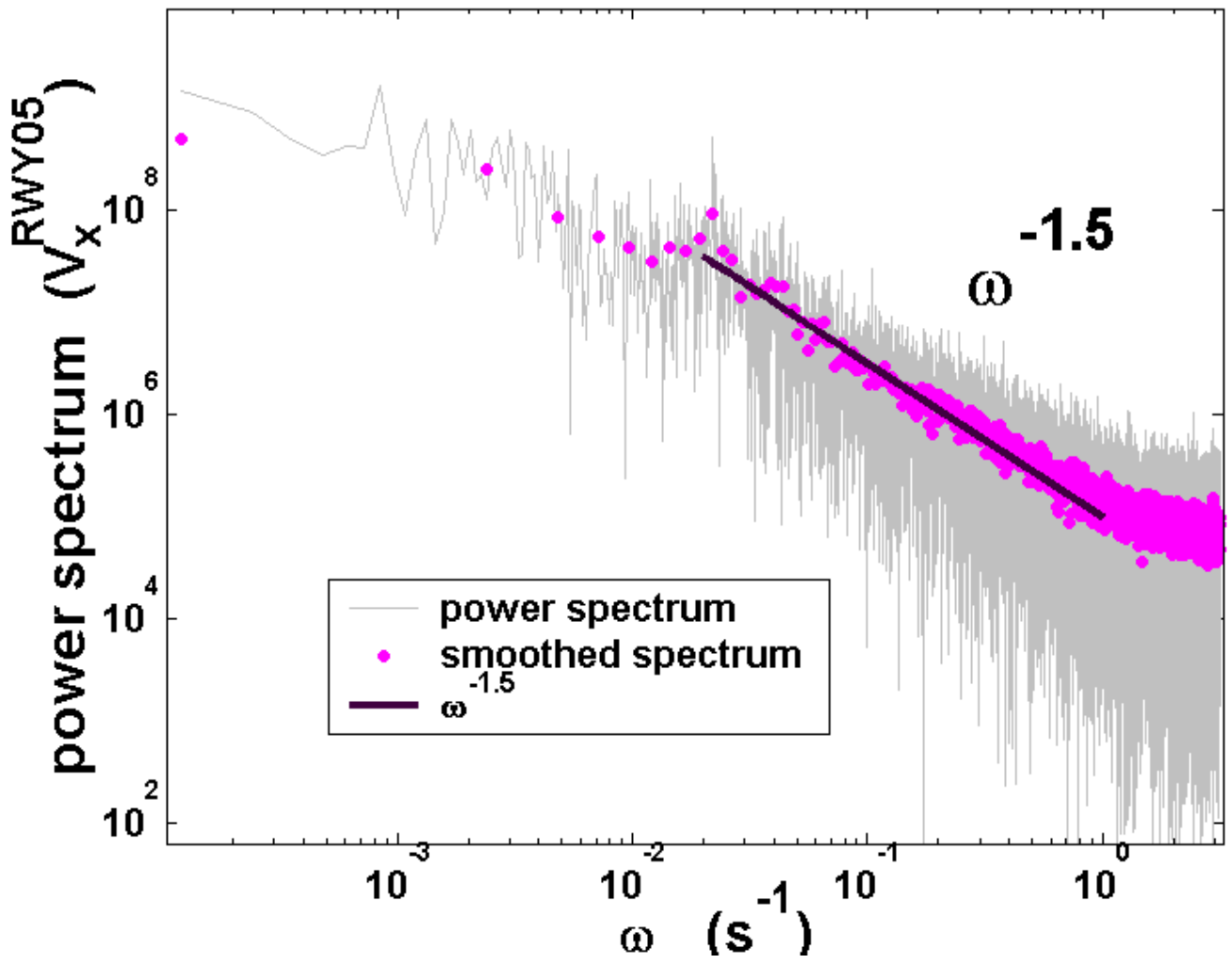

**FIGURE 1.** Power spectrum of the longitudinal component of the wind, recorded by the RWY05 anemometer. The slope we find is -1.5, slightly smaller than that one predicted by Kolmogorov, i.e. -5/3 [8].

## THE ANALYSIS

We have performed an analysis of our time series using the conventional mathematical tools used in small scale physical turbulence. We studied the correlations in the power spectra as well as the probability density functions (pdfs) of the velocity $V$ components returns and of the differences defined as

$$x_k^j(t)_\tau = V_k^j(t+\tau) - V_k^j(t) \quad , \tag{1}$$

$$u_k(t) = V_k^{RWY05}(t) - V_k^{RWY23}(t) \quad , \tag{2}$$

with $j = RWY05,\ RWY23$ , $k = x, y$ , *x=longitudinal, y=transversal.*

The time series presents strong fluctuations of the mean and of the standard deviation $\sigma$ , which disappear increasing the time interval over several months. The study of both auto-correlations and cross-correlations show a slowly decaying behavior, with an initial exponential decay followed by a slower power law decay. The effective relaxation time is about 24 hours. Here we do not report this analysis for brevity, but a complete discussion can be found in [7].

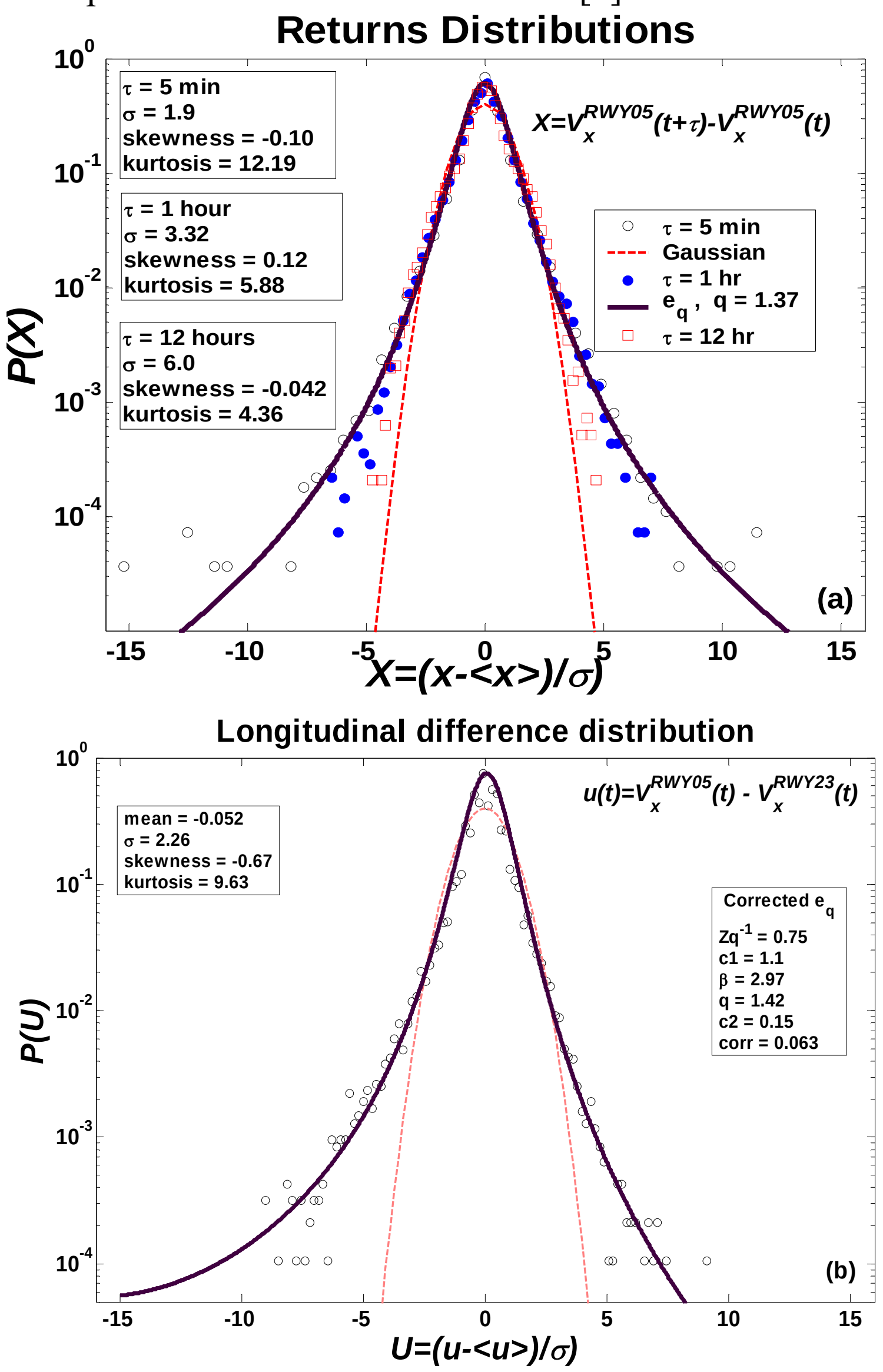

**FIGURE 2 .** (a) Experimental data (symbols) normalized to unit variance are plotted in comparison with a Gaussian curve (dashed curve) and a q-exponential curve (3) with q=1.37 (full line). The index q has been derived from data windowing (1 hr intervals) see text. Considering a window of 12 hr (open squares) the pdf is closer to the Gaussian pdf. (b) Longitudinal components differences (open circles) in comparison with the Gaussian curve (dashed curve) and the q-exponential curve (4) (full curve) with the index q=1.42, extracted from the data. The curve has been corrected for the asymmetry, see text.

Correlations are evident also in the power spectra as shown in Fig. 1, where we plot that one of the longitudinal component of the anemometer RWY05 . We found a slope, for the high-mid frequency part of the spectrum, very similar to the usual Kolmogorov one -5/3 expected for turbulence [8].

By means of the velocity returns (1) and the differences (2), we have calculated the probability density functions (pdfs) which are plotted in Fig. 2. The non-Gaussian features of both velocity returns (a) and the longitudinal velocity differences pdfs (b) is evident in the figure. For comparison we report also the corresponding Gaussian curve normalized to unit variance (dashed curve) and the q-exponential curve (full curve) reported below as eq. (3), with q=1.37, which reproduces very well the experimental data. The velocity differences data are quite similar to the returns ones, although the pdfs of velocity returns are almost symmetric, while for the velocity differences, there is an evident skewness. This asymmetry can be handled correcting the q-exponential, see eq. (4) below, but the entropic index is quite similar, i.e. q=1.42.

With the aim of reproducing the fat tails of velocity pdfs due to the intermittency typical of turbulent phenomena [8], we studied the distribution of the fluctuations of the standard deviation of our data, considering the technique of the moving time window.

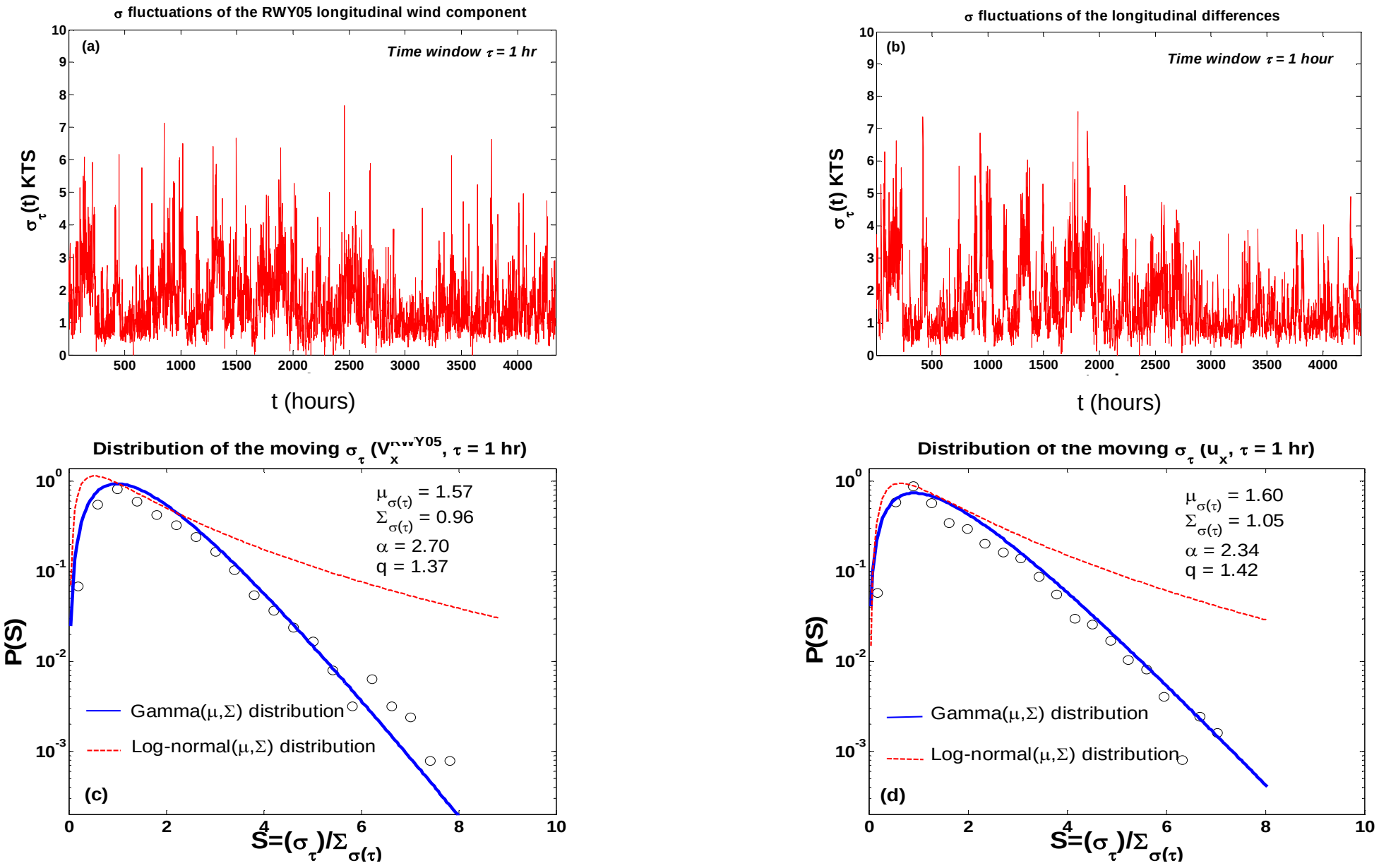

**FIGURE 3.** (a) Fluctuations of the standard deviation of the longitudinal wind component recorded by anemometer RWY05. (b) The same as (a) for the longitudinal wind difference. (c) Standardized distribution (open dots) of fluctuations in (a). (d) The same as in (c) for the fluctuations reported in (b). The Gamma and the Log-normal distribution have been obtained directly from the data considering that they share same average $\mu$ and variance $\Sigma$. Then one can obtain the corresponding nonextensive entropic index $q$ of the Tsallis distribution, see text. For this analysis we used a time window of 1 hr.

The fluctuations (calculated over a moving window of 1 hr) of the standard deviation in both components and differences are shown in Fig.3 (a),(b). The analysis does not depend in a crucial way on the time window as far as the time interval is smaller than the correlation time. The experimental distribution of fluctuations, reported as open circles in Fig.3(c),(d) are well reproduced by a Gamma distribution (full curve) obtained considering the average and standard deviation taken from the data. A tentative fit with a Log-normal distribution, often used in micro-scale turbulence fails completely in reproducing the experimental distribution, see dashed curve in Fig.3 (c),(d). Then, following the *superstatistics* approach introduced by Beck and Cohen [5] and considering the Gamma distribution which reproduces the standard deviation fluctuations, one gets exactly the Tsallis q-exponential pdfs, plotted in Fig.2, which successfully reproduce the experimental velocity pdfs.

The term superstatistics, stands for a statistics of the statistics. It has been proved in [5], that if the distribution of the fluctuations of the standard deviation $\sigma_\tau$ of a fluctuating variable $x$ (velocity returns or differences in our case) is a Gamma distribution, i.e. $f(\sigma_\tau) = \frac{1}{\Gamma(\alpha)}\left(\frac{\alpha}{\mu}\right)^{\alpha} \sigma_\tau^{\alpha-1} e^{-\frac{\alpha\sigma}{\mu}}$ with $\alpha = \left(\frac{\mu_{\sigma_\tau}}{\Sigma_{\sigma_\tau}}\right)^2$, and $\mu_{\sigma_\tau}$ , $\Sigma_{\sigma_\tau}$ being the average and the standard deviation of $\sigma_\tau$, extracted from the data, one gets exactly for the pdf of $x$ the generalized Boltzmann-Gibbs weight of nonextensive statistical mechanics introduced by Tsallis [6], i.e. the so-called q-exponential curve

$$P(x) = \left[1 + (q-1)\mu_{\sigma_\tau} \frac{x^2}{2}\right]^{-1/(q-1)} , \tag{3}$$

with $q = (1+\alpha)/\alpha$. Gamma distribution arises very naturally for a fluctuating environment with an effective finite number of degrees of freedom [5]. In our case, the origin of this distribution is not completely clear and could be originated by the nonstationarity of the data. However, disregarding for the moment the theoretical motivations of our velocity fluctuations, the Gamma distribution justifies the experimental pdfs of both returns (1) and with some corrections, also of the velocity differences (2), as shown in Fig. 2(b). By means of the entropic indexes $q$ extracted from the data, i.e. $q = 1.37$ for the returns and $q = 1.42$ for the longitudinal velocity differences, we are able to reproduce the experimental pdfs in a coherent way. For physical turbulence in the inertial range, the intermittency of velocity differences, was justified in [9], [10], by the hypotheses that the energy dissipation rate $\varepsilon_r$ , averaged on a scale r, (or the energy transfer rate as stressed later by Kraichman [11] and Castaing et al. [12] ), is log-normally distributed. However a Gamma distribution model, for turbulent frequency, was developed by Jayesh and Pope [13]. Finally, we notice that in our experiment the Taylor's hypotheses of frozen turbulence seems not applicable, likely for the low-frequency nature of the wind sampling, since the velocity returns (1) cannot be converted into the two-point velocity differences (2).

For what concerns the velocity difference pdfs, in order to correct for the skewness of the experimental pdf, we used the modified q-exponential curve proposed by Beck in [1], with a further slight modification, i.e.

$$P(u)=Z_q^{-1}\left\{1+c_1\left(\frac{2}{5-3q}\right)(q-1)\left[\frac{1}{2}u^2-c_2(q-1)\left(u-\frac{1}{3}u^3\right)\right]\right\}^{\frac{-1}{q-1}}, \qquad (4)$$

where $Z_q$ is a normalization factor, while $c_1$, $c_2$ are constants to be determined. In our case, see Fig. 2(b), we get a very good reproduction of the asymmetry choosing $c_1 = 1.1$ and $c_2 = 0.15$. The detailed discussion of our expression can be found in ref.[7].

## CONCLUSIONS

Performing an analysis of low frequency wind velocity data, measured for six months at Florence airport, we have found surprising affinities between our data and those of canonical small-scale turbulence, ranging from strong correlations in power spectra, to local (returns) and global (differences) non-Gaussian features. We have obtained a Gamma distribution from the standard deviation fluctuations of the data, which has then been used to extract the Tsallis entropic index q and successfully reproduce the wind velocity pdfs within a coherent framework. Further analogies and differences with micro-scale turbulence experiments will be discussed in [7].

## ACKNOWLEDGMENTS

We Thank E.N.A.V. S.p.A. – U.A.A.V. Firenze, for the possibility to study wind velocity data taken at the Florence airport. We thank S. Ruffo for his important help in the analysis of the time series and for the critical reading of this paper. We thank C. Tsallis for initial encouraging and stimulating suggestions. Useful discussions with C. Beck, V. Latora, A. Pluchino and H. Swinney are also acknowledged.